\documentclass[prc,aps,twocolumn,showpacs,amssymb,superscriptaddress,fleqn]{revtex4-1}

\usepackage{amsmath}
\usepackage{color}
\usepackage[table]{xcolor}
\usepackage[english]{babel}
\usepackage{graphicx}
\usepackage{dcolumn}
\usepackage{mathtools}
\usepackage{relsize}
\usepackage{float}
\usepackage{braket}
\usepackage{soul}

\newcommand{\vlwk}{$V_{{\rm low}\mbox{-}k}$}
\newcommand{\thetaeff}{$\Theta_{\rm eff}$}
\newcommand{\thetaeffs}{$\Theta_{\rm eff}$'s}

\newcommand{\zbb}{$0\nu\beta\beta$}
\newcommand{\dbb}{$2\nu\beta\beta$}
\newcommand{\heff}{$H_{\rm eff}$}
\newcommand{\heffs}{$H_{\rm eff}$'s}
\newcommand{\qbox}{$\hat{Q}$~box}
\newcommand{\tbox}{$\hat{\Theta}$~box}
\newcommand{\nme}{$M^{0\nu}$}
\newcommand{\nmel}{$M^{0\nu}_{\rm L}$}

\newcommand{\nmes}{$M^{0\nu}$'s}

\newcommand{\nmeds}{$M^{2\nu}$'s}

\begin{document}

\title{The renormalization of the shell-model neutrinoless
  double-$\beta$ decay operator starting from effective field theory
  at leading order}

\author{L. Coraggio}
\affiliation{Dipartimento di Matematica e Fisica, Universit\`a degli
  Studi della Campania ``Luigi Vanvitelli'', viale Abramo Lincoln 5 -
  I-81100 Caserta, Italy}
\affiliation{Istituto Nazionale di Fisica Nucleare, \\ 
Complesso Universitario di Monte  S. Angelo, Via Cintia - I-80126 Napoli, Italy}
\author{G. De Gregorio}
\affiliation{Dipartimento di Matematica e Fisica, Universit\`a degli
  Studi della Campania ``Luigi Vanvitelli'', viale Abramo Lincoln 5 -
  I-81100 Caserta, Italy}
\affiliation{Istituto Nazionale di Fisica Nucleare, \\ 
Complesso Universitario di Monte  S. Angelo, Via Cintia - I-80126 Napoli, Italy}
\author{S. L. Lyu}
\affiliation{Dipartimento di Matematica e Fisica, Universit\`a degli
  Studi della Campania ``Luigi Vanvitelli'', viale Abramo Lincoln 5 -
  I-81100 Caserta, Italy}
\affiliation{Istituto Nazionale di Fisica Nucleare, \\ 
Complesso Universitario di Monte  S. Angelo, Via Cintia - I-80126 Napoli, Italy}
\author{N. Itaco}
\affiliation{Dipartimento di Matematica e Fisica, Universit\`a degli
  Studi della Campania ``Luigi Vanvitelli'', viale Abramo Lincoln 5 -
  I-81100 Caserta, Italy}
\affiliation{Istituto Nazionale di Fisica Nucleare, \\ 
Complesso Universitario di Monte  S. Angelo, Via Cintia - I-80126 Napoli, Italy}

\begin{abstract}
In this work, we approach for the first time the task to perform a
shell-model calculation of the matrix element for the neutrinoless
double-$\beta$ decay, within a fully-consistent framework where the
expressions of the nuclear Hamiltonian and of the decay operators have
been derived through chiral perturbation theory.
More precisely, the effective shell-model Hamiltonian and all
transition operators have been constructed by way of the many-body
perturbation theory, and then employed to calculate both spectroscopic
properties of the nuclei involved in the decays under our
consideration –- namely $^{48}$Ca, $^{76}$Ge, and $^{82}$Se --, as well
as the nuclear matrix elements of the electromagnetic and neutrinoless
double-$\beta$ decays.
We also present a study of the convergence properties of the
calculated matrix elements in order to provide the elements for an
estimate of the theoretical uncertainty.
\end{abstract}

\pacs{21.60.Cs, 21.30.Fe, 23.40.-s, 27.40.+z, 27.50.+e}

\maketitle
\section{Introduction}\label{intro}
During the first decades of this century, the nuclear structure
community has seen the rise and consolidation of an innovative
approach to the description of atomic nuclei, namely the application
of the principles of effective field theory (EFT) to the dynamics
ruling the interaction of nucleons in the nuclear environment.

The seeds of such a revolution were planted in a few papers by Steven
Weinberg, where it was suggested that nuclei, as low-energy systems,
could be studied in terms of pions and nucleons within the framework
of EFT \cite{Weinberg79,Weinberg90,Weinberg91}.
In such an approach, the long-range component of the nuclear force
should be ruled by the symmetries of low-energy QCD – and in
particular the spontaneously broken chiral symmetry -- and the
short-range dynamics is taken into account by a complete set of
contact terms, which are proportional to low-energy constants (LECs)
that have to be fitted to data.

Chiral perturbation theory (ChPT) has then provided a powerful tool to
construct nuclear Hamiltonians with a direct link to the underlying
theory for the strong force among hadrons (QCD), and to expand nuclear
forces in a consistent structure of two- and many-body components
\cite{Epelbaum09,Machleidt11}.

Moreover, also electroweak currents can be consistently constructed
through ChPT, since this framework accounts for the composite nature
of hadrons through the symmetries of QCD.
As a matter of fact, during the last decade the ChPT expansion of the
Gamow-Teller (GT) decay operator has been applied to study standard
$\beta$-decay processes in different mass regions
\cite{King20,Baroni21,Gnech21,Gnech22,King23,Gysbers19,Coraggio24a},
and has contributed substantially to the understanding of the nature
of the well-known problem of the quenching of the axial coupling
constant $g_A$.

In recent years, EFT has been also applied to construct electroweak
currents corresponding to the neutrinoless double-$\beta$ (\zbb) 
decay, performing a ChPT expansion of the two-body current up to
next-to-next-to-leading order (N$^2$LO) \cite{Cirigliano18a}.
A relevant feature of the derivation of the \zbb~ decay operator
within EFT – within the light Majorana-neutrino exchange scenario --
is the emergence at the leading order (LO) of a contact operator, that
is required to guarantee the renormalizability of the process
\cite{Cirigliano18a,Cirigliano19}.
Such a short-range component does not appear in the standard
expression of the \zbb~ operator, that is characterized only by a
long-range component \cite{Tomoda91,Vergados12}, and also introduces a
new LEC that has to be renormalized by evaluating the $nn \rightarrow
pp e^- e^-$ amplitude \cite{Cirigliano21}.

In this paper, we present the results of shell-model (SM) calculations
which aim to calculate the nuclear matrix element of the \zbb~ decay
\nme~ for a few nuclei of interest -- $^{48}$Ca, $^{76}$Ca, and
$^{82}$Se –, in terms of effective SM Hamiltonians and decay operators
that have been derived from nuclear potentials and \zbb~ currents
which have been constructed through ChPT.

Our present work is a natural extension of a previous one
\cite{Coraggio24a}, where we investigated GT transitions involving the
same nuclear systems to assess the role of both two-body electroweak
currents and many-body correlations as the causes of the well-known
issue of the quenching of the axial coupling constant $g_A$
\cite{Suhonen17b}.
In fact, as in Ref. \cite{Coraggio24a}, we employ effective SM
Hamiltonians (\heffs) derived by way of the many-body perturbation
theory \cite{Kuo81,Suzuki95,Coraggio12a,Coraggio20c}, starting from
the well-known two-nucleon potential constructed by Entem and
Machleidt through a chiral perturbative expansion up to N$^3$LO
\cite{Entem03,Machleidt11}, and paired by a three-body component that
is consistently built at N$^2$LO in ChPT.

Consistently with the construction of \heffs, we derive effective SM
decay operators from two-body current operators at LO in ChPT, by way
of the same many-body approach we have followed in our previous study
\cite{Coraggio20a}, where the calculations were performed starting
from the high-precision CD-Bonn nucleon-nucleon (2NF) potential
\cite{Machleidt01b}, whose repulsive high-momentum components were
renormalized through the \vlwk~ procedure \cite{Bogner02}.
We also confront the outcome of our study with the results reported in
a recent paper of Castillo {\it et al.}, where the authors performed
the calculation of \zbb~ nuclear matrix elements by way of the nuclear
shell model and proton-neutron quasiparticle random-phase
approximation (pnQRPA), using two-body decay currents up to N$^2$LO in
ChPT, but employing phenomenological \heffs~ \cite{Castillo25}.

This work is organized as follows.
In the upcoming Section \ref{theory}, we report a few details about
the expression of the \zbb~ decay operator according to the ChPT
expansion \cite{Cirigliano18a}, and highlight the procedure we have
followed to derive the effective SM Hamiltonian as well as the \zbb~
operator, starting from the chiral 2NF N$^3$LO potential constructed
by Entem and Machleidt \cite{Entem03} and juxtaposed by a three-body
component at N$^2$LO in ChPT \cite{Machleidt11}.

The results of the SM calculations are then presented in Sec.
\ref{results}, where first we report the results of low-energy
spectroscopic properties of the nuclei involved in the \zbb~ decay of
$^{48}$Ca, $^{76}$Ge, and $^{82}$Se, to support the reliability of our
approach to the SM calculation of electroweak transitions by way of
effective Hamiltonians and operators, and validate the nuclear wave
functions of the initial and final states.
Then, we report the calculated nuclear matrix elements \nmes, as well
as a study of the perturbative behavior of those theoretical values to
estimate the theoretical uncertainties that originate from our
approach.

The conclusions of this study are drawn in Sec. \ref{conclusions},
together with the perspectives of our current project.

\section{Theoretical framework}\label{theory}
\subsection{ChPT Hamiltonian and \zbb~ decay operators} \label{theoryho}
As mentioned in the Introduction, during the end of 1990s and the dawn
of 2000s the application of EFT to low-energy nuclear systems,
especially within the ChPT approach, has provided a valuable tool to
tackle the problem of generating hadronic interactions in a low-energy
regime \cite{Epelbaum09,Machleidt11}.
The relevant feature of pursuing this approach to the construction of
nuclear forces is the link with their underlying theory – the quantum
chromodynamics --, that is obtained through the fundamental
requirement that any nuclear effective Hamiltonian has to obey to the
relevant symmetries of QCD \cite{Weinberg79}.

This framework is based on identifying a separation of the scales
between hadronic systems that are characterized by well-defined energy
regimes \cite{vanKolck99}.
For finite nuclei the so-called ``hard scale’’ is set at $\Lambda_\chi
\sim m_\rho \sim 1$ GeV and the soft scale is then identified with the
pion mass --   $Q \sim m_\pi$ --, known also as the chiral-symmetry
breaking scale.
This represents the building block of the low-energy ChPT expansion,
which is arranged in terms of the soft scale over the hard scale,
$(Q/\Lambda_\chi)^\nu$,  $Q$ being an external momentum or a pion
mass, while the degrees of freedom of the nuclear Hamiltonian are
pions and nucleons and, in some cases, ($\Delta$) resonances.

As previously reported, for the present study, we have considered a
nuclear Hamiltonian consisting of a two- (2NF) and a three-nucleon
(3NF) component of the nuclear force.
As regards the 2NF component, we choose the high-precision 
potential developed by Entem and Machleidt through a perturbative
expansion at N$^3$LO \cite{Entem03}, that is characterized by a
regulator function whose cutoff parameter is $\Lambda = 500$ MeV.
This potential exhibits a smooth behavior in the high-momentum regime
and can be profitably employed for a perturbative derivation of
\heff~, as was shown in Refs. \cite{Coraggio12a,Ma19}.
Consistently, the 2NF potential is paired by a 3NF one, derived at
N$^2$LO in ChPT, and consisting of three topologies: the two-pion
exchange (2PE), one-pion exchange (1PE), and three-nucleon-contact
interactions \cite{Machleidt11}.
These terms are specified by a set of LECs that already appear in the
2NF components, but they also contain a new LEC $c_D$ -- associated to
the 1PE contribution --, while another new one, $c_E$ , enters in the
3NF contact potential.
These LECs, $c_D$ and $c_E$ , should be fixed to reproduce the
observables of the $A=3$ system, and we have adopted the same values
as in Refs. \cite{Fukui18,Ma19,Coraggio20e,Coraggio21,Coraggio24a,Lyu25a},
namely $c_D=-1$ and $c_E=-0.34$.
This choice traces back to a study performed within the no-core shell
model (NCSM), where the authors first constrained the relation of
$c_D$-$c_E$, and then investigated a set of observables in light
$p$-shell nuclei to obtain a second constraint \cite{Navratil07a}.

As pointed out in the Introduction, another key feature of ChPT is to
provide a tool to construct also electroweak currents
consistently with the derivation of the nuclear Hamiltonian, rooting
the structure of these decay operators in the symmetries of the QCD
and also accounting for the composite structure of nucleons and pions
\cite{Park93,Pastore09,Kolling09,Baroni16b,Krebs17,Krebs20}.
This approach has been extensively applied to study GT transitions in
light nuclear systems \cite{King20,Baroni21,Gnech21,Gnech22,King23},
in a few medium-mass nuclei through {\it ab initio} methods
\cite{Gysbers19}, and also by way of the realistic shell model
\cite{Coraggio24a}.
The application of ChPT to the study of $\beta$ decay has represented
a turning point in the understanding of the mechanism of the
renormalization of electroweak operators in nuclear systems, and
overcoming the so-called “quenching problem” of the axial coupling
constant $g_A$ \cite{Arima73,Towner83,Martinez-Pinedo96}.

ChPT has been also employed to construct the expression of the \zbb~
decay operator induced by the light Majorana-neutrino exchange,
shifting the physics of such a process from the energy scale of the
lepton-number violation $\Lambda_{\rm LNV}=1-100$ TeV down the one of
low-energy nuclear systems, namely the chiral-symmetry breaking scale
\cite{Cirigliano18a}.

In recent years, the expansion of two- and three-body currents of the
\zbb~ operator has been carried out in ChPT
\cite{Cirigliano18a,Cirigliano18b,Chambers-Wall26}, and a few studies
on the impact of this novel approach to the definition of the \zbb~
operator has been already conducted by calculating the nuclear matrix
elements \nmes~ for nuclei that are currently of experimental interest
\cite{Wirth21,Jokiniemi21,Belley24,Castillo25}.

As a matter of fact, one of the most relevant features of the
definition of the \zbb~ decay operator in ChPT is the emergence of a
short-range contact operator which is not present in the standard
formulation of the theory of \zbb~ decay \cite{Tomoda91,Vergados12}.
In the earliest formulation of the EFT approach to defining the
light-neutrino exchange mechanism of the \zbb~ decay, this contact
term appeared as a contribution at N$^2$LO of the ChPT expansion
\cite{Cirigliano18a}.
Then, in a following study it has been shown that this contribution
had to be promoted at LO of the expansion, to accomplish the need to
introduce a counterterm to absorb the ultraviolet (UV) divergences
that appear in the calculation of the amplitude of the $nn \rightarrow
pp e^-e^-$ process for $^1S_0 \rightarrow ^1S_0$ transitions
\cite{Cirigliano19}.

As a counterterm, this contribution is regulated by a LEC, dubbed
$g_{\nu}^{\rm NN}$, and the only way to fix it would be by reproducing
the data of lepton-number-violation (LNV) processes, data that are not
available at present.

In Ref. \cite{Cirigliano19}, the authors suggested to estimate the
magnitude of the contact term by considering the coefficients of the
charge-independence breaking (CIB) contact interaction.
For the Entem-Machleidt N$^3$LO potential this leads to $g_{\nu}^{\rm
  NN}=-0.47$ fm$^2$, that is the value we have employed in this work.
It is worth pointing out that Jokiniemi {\it et al.} have followed the
same considerations in Ref. \cite{Jokiniemi21}.

As can be seen in the following, these considerations about the
contact term, that is intimately connected with the renormalization
mechanism of ChPT approach, are very important in the definition of
the nuclear matrix element \nme.

It should be recalled that the expression of the half-life of a \zbb~
decay, assuming the exchange of a light Majorana neutrino, is:

\begin{equation}
\left[ T^{0\nu}_{1/2}\right]^{-1} = G^{0\nu} g_A^4 \left| M^{0\nu} \right|^2
\left|\frac{ \langle m_{\nu}\rangle}{m_e}\right|^2~,
\label{halflife}
\end{equation}

\noindent
where $G^{0\nu}$ is the phase-space factor \cite{Kotila12,Kotila13},
\nme~ is the nuclear matrix element directly related to the wave
functions of the parent and grand-daughter nuclei, $g_A$ is the axial
coupling constant, $m_e$ is the electron mass, and $\langle m _{\nu}
\rangle = \sum_i (U_{ei})^2 m_i$ is the effective neutrino mass, as
expressed in terms of the neutrino masses $m_i$ and their mixing
matrix elements $U_{ei}$.

At LO in ChPT, \nme~ is expressed as the sum of a long- and
short-range components \cite{Cirigliano18a}:

\begin{equation}
M^{0\nu} = M^{0\nu}_{\rm L} + M^{0\nu}_{\rm S}
\label{nmeLS}
\end{equation}

In turn, the formal expression of the LO long-range component \nmel~
is expressed in terms of the two-body transition-density matrix
elements $\langle f | a^{\dagger}_{p}a_{n} a^{\dagger}_{p^\prime}
a_{n^\prime} | i \rangle$, between the initial ($i$) and final ($f$)
nuclear wave functions, as:

\begin{eqnarray}
M_{\rm L \alpha }^{0\nu}  & =  & \sum_{j_n j_{n^\prime} j_p j_{p^\prime}}
\langle f | a^{\dagger}_{p}a_{n} a^{\dagger}_{p^\prime} a_{n^\prime} 
| i \rangle \nonumber \\
~ & ~& \times  \left< j_p  j_{p^\prime} \mid \Theta_{\rm L \alpha} \mid  j_n j_{n^\prime}
       \right>~, \label{M0nuappl}
\end{eqnarray}

\noindent
where $\alpha$ stands for Fermi ($F$), Gamow-Teller (GT), or tensor
($T$) decay channels, and the operators $\Theta_{\rm L \alpha}$ are
\cite{Engel17}:

\begin{eqnarray}
 \Theta_{\rm GT} & = & [ \tau^-_{1} \tau^-_{2} (\vec{\sigma}_1 \cdot \vec{\sigma}_2) H_{\rm
GT}(r) ]\label{operatorGT} \, , \\
\Theta_{\rm F} & = & \tau^-_{1} \tau^-_{2} H_{\rm F}(r)
                          \label{operatorF} \, ,\\
\Theta_{\rm T} & = & [\tau^-_{1} \tau^-_{2} \left(
                         3\left(\vec{\sigma}_1 \cdot \hat{r} \right)
                         \left(\vec{\sigma}_2 \cdot \hat{r} \right) -
                         \right. \nonumber \\
  ~ & ~ & \left. \vec{\sigma}_1 \cdot \vec{\sigma}_2 \right) H_{\rm
          T}(r) ]~, \label{operatorT}
\end{eqnarray}

The neutrino potentials are expressed as \cite{Cirigliano18a}:

\begin{equation}
H_{\alpha}(r)=\frac {2R}{\pi} \int_{0}^{\infty} \frac {j_{n_{\alpha}}(qr)
  h_{\alpha}(q^2)qdq}{q}~.
\label{neutpotappl}
\end{equation}

\noindent
It should be pointed out that at LO the expression of \nmel, as well
as the one of the neutrino potentials in Eq. (\ref{neutpotappl}), is the
same as the one in the standard formulation \cite{Tomoda91,Vergados12}
when resorting to the closure approximation \cite{Engel17}, but
setting the closure energy equal to zero \cite{Cirigliano18a}.

Coming back to Eq. (\ref{neutpotappl}), the nuclear radius $R$ is defined
as $R=1.2 A^{1/3}$ fm, $j_{n_{\alpha}}(qr)$ is the spherical Bessel
function, $n_{\alpha}=0$ for Fermi and Gamow-Teller components, while
$n_{\alpha}=2$ for the tensor component.

In the following, we also report the explicit expressions of neutrino
form functions, $h_{\alpha}(q)$, for light-neutrino exchange at LO
\cite{Cirigliano18a}:

\begin{eqnarray}
h_{\rm F} ({ q}^{2})  & = & g^2_V \, ,  \nonumber \\
h_{\rm GT} ({ q}^{2}) & = & 
\left[ 1 - \frac{2}{3} \frac{ { q}^{2}}{ { q}^{2} + m^2_\pi } + 
\frac{1}{3} ( \frac{ { q}^{2}}{ { q}^{2} + m^2_\pi } )^2 \right]
\nonumber\\
&& + \frac{2}{3} \frac{g^2_M}{g^2_A} \frac{{ q}^{2} }{4 m^2_p }, 
\nonumber \\
h_{\rm T} ({ q}^{2}) & = & \left[ 
\frac{2}{3} \frac{ { q}^{2}}{ { q}^{2} + m^2_\pi } -
\frac{1}{3} ( \frac{ { q}^{2}}{ { q}^{2} + m^2_\pi } )^2 \right] 
\nonumber\\
&& + \frac{1}{3} \frac{g^2_M}{g^2_A} \frac{{ q}^{2} }{4 m^2_p }  \, ,   
\end{eqnarray}
\noindent
where $g_V = 1$, $g_A \equiv g_A^{free}=1.2723$, $g_M=(\mu_p -
\mu_n)g_V$, and $(\mu_p - \mu_n) =
4.7$.

Then, the full expression of the long-range term (at LO) of the
nuclear matrix element \nme~ is written as
\begin{equation}
M^{0\nu}_{\rm L} =  M_{\rm GT}^{0\nu} - \frac{g_V^2}{g_A^2}  M_{\rm F}^{0\nu}
+  M_{\rm T}^{0\nu}~~.
\label{nme00nu}
\end{equation}

As regards the short-range contact term, its expression is the same as
in Eq. \ref{M0nuappl}

\begin{eqnarray}
M_{\rm S}^{0\nu} & =  & \sum_{j_n j_{n^\prime} j_p j_{p^\prime}}
\langle f | a^{\dagger}_{p}a_{n} a^{\dagger}_{p^\prime} a_{n^\prime} 
| i \rangle \nonumber \\
~ & ~& \times  \left< j_p  j_{p^\prime} \mid \Theta_{\rm S} \mid  j_n j_{n^\prime}
       \right>~, \label{M0nuappCT}
\end{eqnarray}

\noindent
where $\Theta_{\rm S} $  is:

\begin{eqnarray}
\Theta_{\rm S} & = & \tau^-_{1} \tau^-_{2} H_{\rm S}(r)
                          \label{operatorCT} \, ,
\end{eqnarray}
\noindent 
and the neutrino potential is:

\begin{equation}
H_{\rm S}(r)=\frac {2R}{\pi} \int_{0}^{\infty} j_{0}(qr)
  h_{\rm S}(q^2)q^2dq~,
\label{neutpotappCT}
\end{equation}

\noindent
where we choose to regularize the contact term with a Gaussian
regulator as in Refs. \cite{Jokiniemi21,Castillo25}:

\begin{equation}
h_{\rm S} (q^2) = -2 (g^{\rm NN}_{\nu}/g_A^2) e^{-q^2/(2\Lambda^2)}~.
\label{pots}
\end{equation}

\subsection{Effective SM Hamiltonian and transition operators}\label{effhop}
In this section we are going to sketch out briefly our approach to the
derivation of the effective SM operators that are necessary to
construct the nuclear wave functions – namely, the effective SM
Hamiltonian \heff~--, and then calculate the nuclear matrix elements
that are needed to extract the half-lives of electromagnetic and
$\beta$ decays, the effective SM decay operators \thetaeffs.

A more detailed presentation of our approach to the derivation of
effective SM Hamiltonians and transition operators by way of the
many-body perturbation theory, and starting from realistic nuclear
forces, can be found in
Refs. \cite{Coraggio12a,Coraggio20c,Coraggio24b}.

As mentioned in the Introduction, we construct the two \heffs, for
$0f1p$ and $0f_{5/2}1p0g_{9/2}$ model spaces, respectively, starting
from the  ChPT nuclear Hamiltonian as described in
Sec. \ref{theoryho}.

It is also worth noting that the Coulomb potential is explicitly
included in the proton-proton channel of the nuclear Hamiltonian. 
 
The procedure to derive \heff~ starts from the SM nuclear Hamiltonian
$H$ for $A$ interacting nucleons, which, by introducing an auxiliary
potential $U$, is split into a one-body term $H_0$, whose eigenvectors
provide the unperturbed SM basis, and an interaction component $H_1$:

\begin{eqnarray}
 H &= & T + V_{\rm 2NF} + V_{\rm 3NF}= (T+U)+(V_{\rm 2NF}+ V_{\rm 3NF}-U) \nonumber\\
~& = &H_{0}+H_{1}~,\label{smham}
\end{eqnarray}

\noindent
where the auxiliary potential $U$ is chosen to be the
harmonic-oscillator (HO) one, with a value according to the
Blomqvist-Molinari formula $\hbar \omega = 45 A^{-1/3} - 25 A^{-2/3}$
MeV \cite{Blomqvist68}, namely $\hbar \omega = 11$ and 10 MeV for
$^{40}$Ca and $^{56}$Ni cores, respectively .

Since the eigenvalue problem of $H$ for a many-body system, and within
an infinite Hilbert-space of $H_0$ eigenvectors, cannot be solved,
there is the need to construct an effective Hamiltonian by way of a
similarity transformation \cite{Suzuki80,Stroberg19} to project the
eigenvalue problem into a truncated model space.
In our case, we study the double-$\beta$ decay of $^{48}$Ca
considering the model space by four proton and neutron orbitals
$0f_{7/2}, 0f_{5/2}, 1p_{3/2}, 1p_{1/2}$, outside $^{40}$Ca
doubly-closed core.
As regards the decays of $^{76}$Ge and $^{82}$Se, the model space is
spanned by the four HO orbitals $0f_{5/2}, 1p_{3/2}, 1p_{1/2},
0g_{9/2}$, now considering $^{56}$Ni as reference nucleus.

These are the same model spaces as in Ref. \cite{Coraggio24a}, where
we have performed a study of single- and double-$\beta$ decays of the
same isotopes.

The \heffs~ for these model spaces are derived by way of the
time-dependent perturbation theory, namely performing the
Kuo-Lee-Ratcliff folded-diagram expansion in terms of the
$\hat{Q}$-box vertex function \cite{Kuo90,Hjorth95,Coraggio12a}:

\begin{equation}
H^{\rm eff}_1 (\omega) = \hat{Q}(\epsilon_0) - P H_1 Q \frac{1}{\epsilon_0
  - Q H Q} \omega H^{\rm eff}_1 (\omega) ~,\label{eqfinal}
\end{equation}
\noindent
where $\omega$ is the wave operator decoupling the model space $P$ and
its complement $Q$, and $\epsilon_0$ is the eigenvalue of the unperturbed
degenerate HO Hamiltonian $H_0$.

We recall that \qbox~ operator is defined as
\begin{equation}
\hat{Q} (\epsilon) = P H_1 P + P H_1 Q \frac{1}{\epsilon - Q H Q} Q
H_1 P ~, \label{qbox}
\end{equation}
\noindent
and $\epsilon$ is an energy parameter called the ``starting energy''.

Computationally, it is impossible to perform an exact calculation of
the \qbox, then the term $1/(\epsilon - Q H Q)$ is expanded as a power
series

\begin{equation}
\frac{1}{\epsilon - Q H Q} = \sum_{n=0}^{\infty} \frac{1}{\epsilon -Q
  H_0 Q} \left( \frac{Q H_1 Q}{\epsilon -Q H_0 Q} \right)^{n} ~.
\end{equation}

In all our applications of the many-body perturbation theory to derive
\heff, we expand the \qbox~ up to the third order in perturbation
theory (n=1) \cite{Coraggio20c}, and the set of two-body
configurations belonging to the subspace $Q$ is truncated to those
which correspond to an excitation energy smaller than $E_{\rm max}=
N_{\rm max} \hbar \omega$ \cite{Coraggio12a}.
In the present paper, we employ the same \heffs~ as in
Ref. \cite{Coraggio24a}, where all effective SM operators and
Hamiltonians have been calculated through $N_{\rm max} = 18$.

It should be pointed out that we have performed previous studies
\cite{Coraggio18,Ma19}, where it was verified that such a value of
$N_{\rm max}$ is large enough to obtain convergent values of the
single-particle (SP) energies and two-body matrix elements of the
residual interaction (TBMEs), that is the basic requirement for stable
values of the excitation spectra and transition strengths, with a
fixed decay operator.

After the \qbox~ is calculated perturbatively, the non-linear matrix
equation (\ref{eqfinal}) can be solved by way of iterative techniques
\cite{Krenciglowa74,Suzuki80}, or graphical non-iterative methods
\cite{Suzuki11}.

Usually, SM calculations are performed by a number of valence nucleons
that are larger than two, then we include also contributions from
induced three-body forces in the calculation of the \qbox, namely
involving also three valence nucleons, and then resort to a
normal-ordering decomposition of the 3NF induced-force contributions
arising at second order in perturbation theory.
More details of such a method are reported in
Refs. \cite{Coraggio20c,Coraggio20e}, but it is worth pointing out
that the chosen reference state is the ground state of the nucleus
under study, and we consider a fractional filling of model-space
orbitals, a procedure that is performed also for the application of
the valence-space in-medium similarity transformation group (VS-IMSRG)
approach within the “target” normal ordering \cite{Stroberg17}.

We have adopted the normal-ordering decomposition also calculating the
contributions at first order in many-body perturbation theory for the
calculation of the \qbox~ of the N$^2$LO 3NF component of the ChPT
Hamiltonian \cite{Fukui18}.

The SM parameters of the \heffs~ that have been employed to study the
\zbb~ decay of $^{48}$Ca, $^{76}$Ge, and $^{82}$Se -- namely the SP
energies and the TBMEs of the residual interaction -- are included in
the Supplemental Material in Ref. \cite{Coraggio24a}.

Now, we turn our attention to the derivation of effective SM decay
operators \thetaeffs.

The need to build effective operators is due to the issue that the
diagonalization of the \heff~ does not lead to the true nuclear
wave-functions, but to their projections onto the model space $P$.
This means that, consistently with the construction of \heff, any
decay operator $\Theta$ has to be renormalized, so that \thetaeff~
accounts for the neglected degrees of freedom belonging to the
subspace $Q=1-P$.

As in our previous studies of double-$\beta$ decay processes
\cite{Coraggio17a,Coraggio19a,Coraggio20a,Coraggio22a,Coraggio24a},
the construction of the effective SM \zbb~ operators has been carried
out by way of the approach that has been introduced by Suzuki and
Okamoto \cite{Suzuki95}, that is consistent with the derivation of
\heff.

As for \heff, the derivation of \thetaeff~ is grounded on the
perturbative expansion of a vertex function, the so-called \tbox,
which is the counterpart of the \qbox~ that has been previously
defined in Eq. (\ref{qbox}).
The details of the procedure can be found in
Refs. \cite{Suzuki95,Coraggio20c}, here we are only going to outline
the structure of the derivation of any effective SM decay operators
\thetaeff~ by way of many-body perturbation theory.

As previously mentioned, the perturbative calculation of \thetaeff~
starts from introducing two energy-dependent vertex functions:

\[
\hat{\Theta} (\epsilon) = P \Theta P + P \Theta Q
\frac{1}{\epsilon - Q H Q} Q H_1 P ~, \]
\[ \hat{\Theta} (\epsilon_1 ; \epsilon_2) = P H_1 Q
\frac{1}{\epsilon_1 - Q H Q} Q \Theta Q \frac{1}{\epsilon_2 - Q H Q} Q H_1 P ~,\]

\noindent
and of their derivatives calculated in $\epsilon=\epsilon_0$,
$\epsilon_0$ being the eigenvalue of the degenerate unperturbed
Hamiltonian $H_0$:

\[
\hat{\Theta}_m = \frac {1}{m!} \frac {d^m \hat{\Theta}
 (\epsilon)}{d \epsilon^m} \biggl|_{\epsilon=\epsilon_0} ~, \]
\[ \hat{\Theta}_{mn} =  \frac {1}{m! n!} \frac{d^m}{d \epsilon_1^m}
\frac{d^n}{d \epsilon_2^n}  \hat{\Theta} (\epsilon_1 ;\epsilon_2)
\biggl|_{\epsilon_1= \epsilon_0, \epsilon_2  = \epsilon_0} ~\]

Then, a series of operators $\chi_n$ is calculated:

\begin{eqnarray}
\chi_0 &=& (\hat{\Theta}_0 + h.c.)+ \hat{\Theta}_{00}~~,  \label{chi0} \\
\chi_1 &=& (\hat{\Theta}_1\hat{Q} + h.c.) + (\hat{\Theta}_{01}\hat{Q}
+ h.c.) ~~, \label{chi1} \\
\chi_2 &=& (\hat{\Theta}_1\hat{Q}_1 \hat{Q}+ h.c.) +
(\hat{\Theta}_{2}\hat{Q}\hat{Q} + h.c.) + \nonumber \\
~ & ~ & (\hat{\Theta}_{02}\hat{Q}\hat{Q} + h.c.)+  \hat{Q}
\hat{\Theta}_{11} \hat{Q}~~, \label{chi2} \\
&~~~& \cdots \nonumber
\end{eqnarray}

\noindent
where $\hat{Q} \equiv \hat{Q}(\epsilon)$.

Finally, \thetaeff~ is expressed in the following form:
\begin{equation}
\Theta_{\rm eff} = H_{\rm eff} \hat{Q}^{-1}  (\chi_0+ \chi_1 + \chi_2 +\cdots) ~,
\label{effopexp}
\end{equation}

\noindent
where the $\hat{\Theta}$ function is expanded up to third order in
perturbation theory, consistently with the perturbative calculation of
the \qbox.

It should be noticed that, since the \zbb-decay operator owns a
two-body structure and we are considering nuclear systems with a
number of valence nucleons far larger than two, we have included in
the \tbox~ expansion also leading-order three-body contributions,
namely the second-order three-body diagrams that can be found in
Refs. \cite{Coraggio20a,Coraggio20c}.
In Ref. \cite{Coraggio20a} we have also discussed the impact of these
three-body contributions to the definition of \thetaeff~ for the \zbb~
decay.

In Refs. \cite{Coraggio18,Coraggio19a,Coraggio20a} there are reported
studies of the convergence of the $\chi_n$ series and of the
perturbative properties of the \tbox, in order to support the
robustness of the expansion of \thetaeff.

It is worth pointing out that one of the themes of discussion of the
results in Sec. \ref{results} will be the evaluation of theoretical
uncertainties associated to our calculated \nmes~ of the \zbb~ decays
of $^{48}$Ca, $^{76}$Ge, and $^{82}$Se, uncertainties that originate
from the perturbative expansion of \heff~ and \thetaeff.

\section{Results}
\label{results}
In this section we present the results of our SM calculations, that
are obtained by employing \heffs~ and \thetaeffs~ from ChPT.

First, it is worth considering the quality of the agreement between
calculated and experimental spectroscopic properties of the nuclei
which are the focus of our study, aiming to assess the quality of the
nuclear wave functions we employ to calculate \nmes.
The results of such a study have already been reported in
Ref. \cite{Coraggio24a}, and in Sec. \ref{spectro} we will report a
study of the perturbative properties of \heffs~ and decay \thetaeffs.

Then, in Sec. \ref{nmeLO} we report the results of the calculation of
\nmes~ for $^{48}$Ca, $^{76}$Ge, and $^{82}$Se decays, considering the
\zbb~ decay operator at the LO in ChPT expansion.
We will present an analysis of the perturbative properties of the
\zbb~ \thetaeff~ to ascertain an uncertainty estimate of the
calculated \nmes, and compare the results with those we obtained in a
similar study, in which we have derived \heffs~ and \thetaeffs~ from
the meson-theoretic CD-Bonn 2NF potential \cite{Coraggio20a}.

\subsection{Theoretical results vs. experimental quantities}\label{spectro}
As already pointed out in the previous sections, our SM calculations are
carried out employing theoretical SP energies, TBMEs, and effective
transition operators as reported in Sec. \ref{effhop}, whose details
can be found in Ref. \cite{Coraggio20c}.

\begin{center}
\begin{figure}[ht]
\includegraphics[scale=0.45,angle=0]{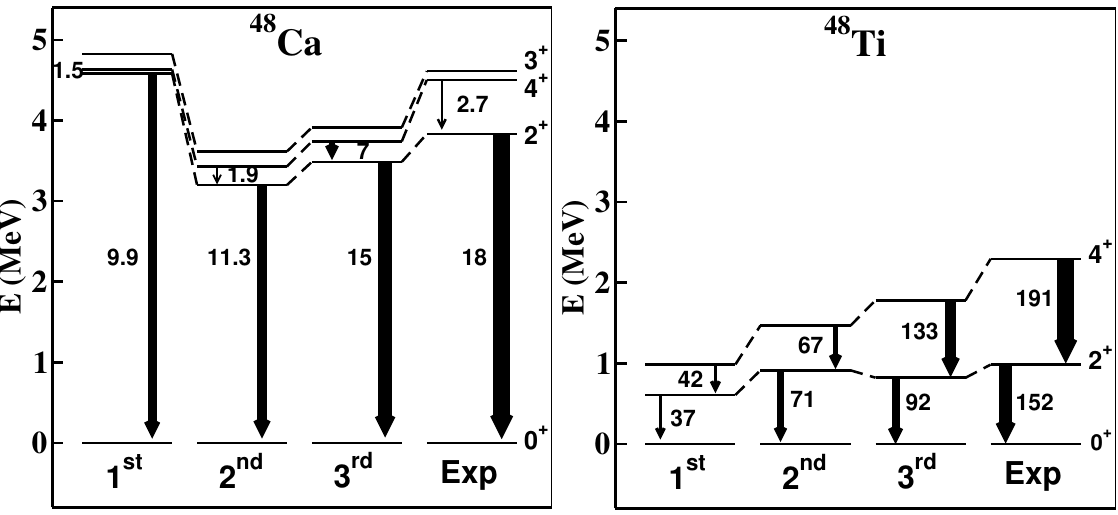}
\caption{Experimental and calculated spectra of $^{48}$Ca and
  $^{48}$Ti. $B(E2)$ strengths (in $e^2{\rm fm}^4$) are also reported
  (see text for details).}
\label{48Ca48Ti}
\end{figure}
\end{center}

We start considering the nuclei involved in the double-$\beta$ decay
of $^{48}$Ca, namely the latter and $^{48}$Ti.
In Fig. \ref{48Ca48Ti}, we show the experimental \cite{ensdf} and
calculated low-energy spectra of both nuclei, obtained within the
full $fp$ shell, namely the proton and neutron $0f_{7/2}$, $0f_{5/2}$,
$1p_{3/2}$, and $1p_{1/2}$ orbitals.
The arrow widths are proportional to the $B(E2)$ strengths, and next
to them we report also the theoretical and experimental values in
$e^2$fm$^4$ \cite{ensdf}.

Since we want to evaluate the perturbative behavior of the calculated
SM effective Hamiltonian, we compare the low-lying excitation spectra
obtained with \heffs~ derived at first-, second-, and third-order in
many-body perturbation theory.
A similar study, performed for one- and two-valence nuclei, was
reported in Ref. \cite{Ma19}, and here we want to extend it to
many-valence nucleon systems.

It should be noticed that in the present study we have employed the same
effective $E2$ transition operator, the one derived at third order in
perturbation theory, to calculate $B(E2)$'s with \heffs~ that are
calculated at different order.
In such a way, the focus is spotted on the quality of the shell-model
wave functions at each order in perturbation theory, so to recover
these information when the attention will be turned to the calculation
of \zbb~ matrix elements.

As can be seen, the observed shell closure of the neutron $0f_{7/2}$
orbital in $^{48}$Ca is reproduced at all orders, a feature that
traces back to the contribution of the three-body component of the
nuclear Hamiltonian \cite{Ma19}.
The perturbative behavior of the results obtained with second- and
third-order \heffs~ is satisfactory, as well as the agreement with
experiment.
First-order results exhibit some overestimate of the shell closure,
that is testified also by the smaller $B(E2; 2^+_1 \rightarrow 0^+_1)$
with respect to second- and third-order ones.

\begin{center}
\begin{figure}[H]
\includegraphics[scale=0.45,angle=0]{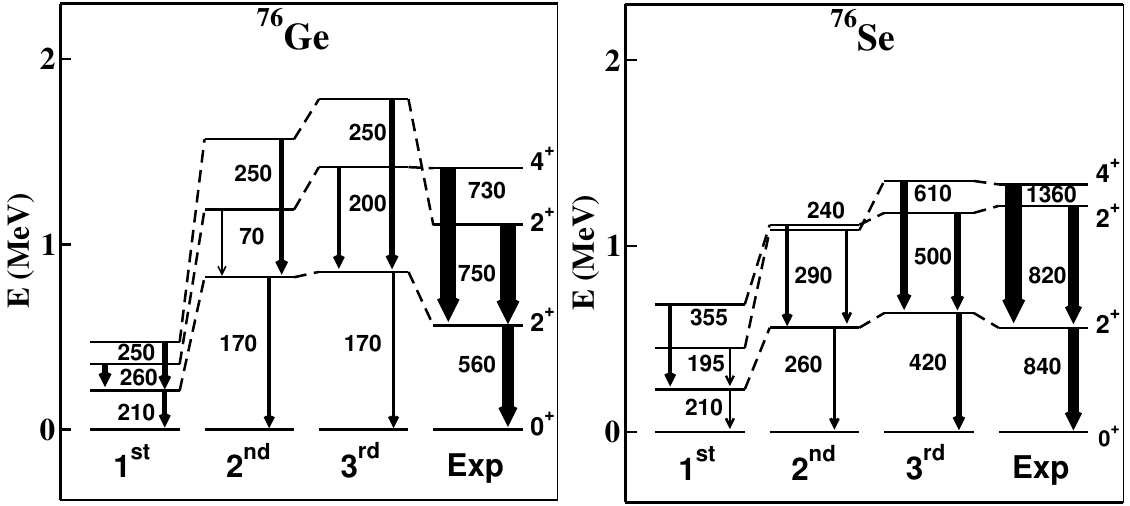}
\caption{ Same as in Fig. \ref{48Ca48Ti}, but for $^{76}$Ge and
  $^{76}$Se (see text for details).}
\label{76Ge76Se}
\end{figure}
\end{center}

As regards $^{48}$Ti, the strength of the proton-neutron interaction
induces a larger collectivity in its low-energy excitation spectrum,
also evidenced by larger $B(E2)$'s.
Second- and third-order \heffs~ reproduce quite well these
experimental features, while first-order \heff~ is characterized by
smaller $B(E2)$'s, despite the compressed energy spectrum which usually
leads to a larger collectivity.

The calculation of low-energy spectra and wave functions of
$^{76}$Ge, $^{76}$Se, $^{82}$Se, and $^{82}$Kr has been carried out by
employing the four proton and neutron orbitals $0f_{5/2}$, $1p_{3/2}$,
$1p_{1/2}$ and $0g_{9/2}$, outside $^{56}$Ni closed core, as model
space.

Fig. \ref{76Ge76Se} reports the experimental \cite{ensdf} and
calculated low-energy spectra of $^{76}$Ge an  $^{76}$Se, as well as
the electric quadrupole transition strengths, following the same
scheme of analysis as for $^{48}$Ca and $^{48}$Ti.
It is worth pointing out that the reproduction of the observables
characterizing low-energy states in these nuclides is far more
challenging for a microscopic nuclear structure model, owing to the
experimental evidence of the $^{76}$Ge rigid triaxial deformation
\cite{Toh13}.

First, we observe that there is quite a satisfactory perturbative
behavior between second- and third-order \heffs~, as regards the
spectroscopic properties of low-energy states.
As regards the agreement with experiment, we have already noticed in
Ref. \cite{Coraggio24a} that the agreement between the experimental
and calculated spectra and $B(E2)$’s for $^{76}$Se is very
satisfactory, but this is not the same for $^{76}$Ge.
We have also shown that this traces back to the monopole component of
the \heff, especially when comparing with the results obtained in our
earlier study where \heff~ was derived from a \vlwk~potential obtained
from the CD-Bonn potential \cite{Coraggio19a}.

\begin{center}
\begin{figure}[H]
\includegraphics[scale=0.45,angle=0]{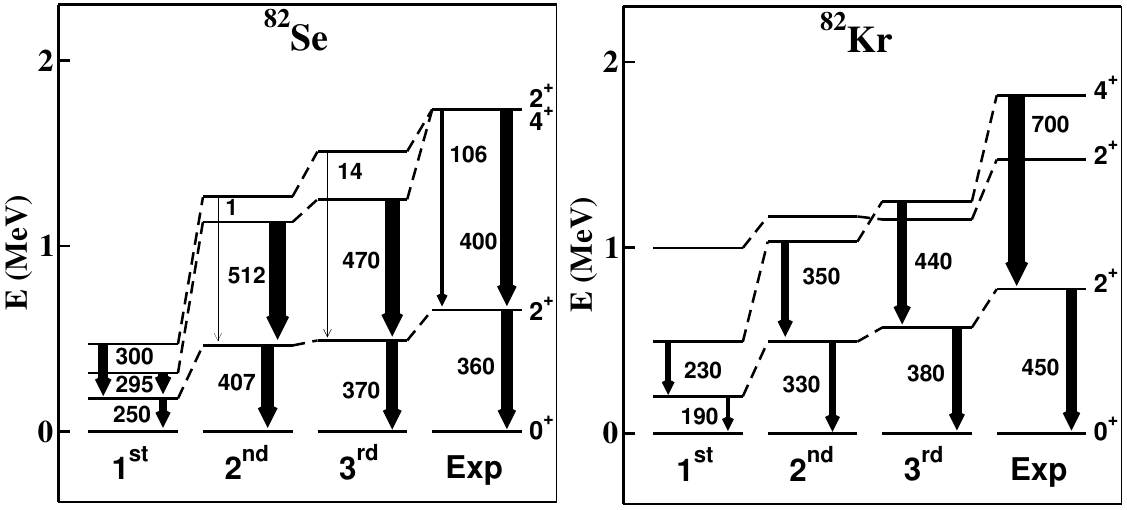}
\caption{ Same as in Fig. \ref{48Ca48Ti}, but for $^{82}$Se and
  $^{82}$Kr (see text for details).}
\label{82Se82Kr}
\end{figure}
\end{center}

In Fig. \ref{82Se82Kr} we report the results for $^{82}$Se and
$^{82}$Kr low-energy spectra, and compare them with the experimental
ones.
As can be seen, we have obtained both a good perturbative behavior of
the \heffs, and a quantitative agreement with observables.

Now, we shift our attention to the results of the matrix elements
\nmeds~ for the two-neutrino double-$\beta$ decay of $^{48}$Ca,
$^{76}$Ge, and $^{82}$Se.
The SM effective operator has been constructed by way of ChPT, namely
the one- and two-body matrix elements of the axial currents ${\mathbf
  J_A}$ are derived through a chiral expansion up to N$^3$LO, and the
LECs appearing in their expression are consistent with those of the
nuclear potential.
The details about the derivation of the SM effective decay operator
are reported in Ref. \cite{Coraggio24a}, here we have employed the
same one – calculated at third order in many-body perturbation theory
--, but diagonalizing the SM Hamiltonian with first-, second-, and
third-order \heffs.

As for the calculation of the $B(E2)$'s, our intent is to focus to the
perturbative behavior of the SM wave functions, as well as their
reliability to reproduce data, within the perspective of the
prediction of \zbb~ nuclear matrix elements.

\begin{table}[ht]
  \caption{Experimental \cite{Barabash20} and calculated \nmeds~(in
    MeV$^{-1}$) for $^{48}$Ca, $^{76}$Ge, and $^{82}$Se \dbb~ decay
    (ground state to ground state).}
\begin{ruledtabular}
\begin{tabular}{ccccc}
\label{ME_2nbb}
  Decay & 1st order & 2nd order & 3rd order & Expt \\
  ~ & ~ & ~ & ~ & ~ \\
 $^{48}$Ca$\rightarrow$$^{48}$Ti & 0.055 & 0.014 & 0.019 & $0.042 \pm 0.004$ \\
 $^{76}$Ge$\rightarrow$$^{76}$Se & 0.022 & 0.056 & 0.118 &$0.129 \pm 0.004$ \\
 $^{82}$Se$\rightarrow$$^{82}$Kr & 0.042 & 0.099 & 0.095 & $0.103 \pm 0.001$ \\
\end{tabular}
\end{ruledtabular}
\end{table}

Table \ref{ME_2nbb} reports the calculated \nmeds, obtained with the
different \heffs, for the \dbb~ decays
$^{48}$Ca$\rightarrow$$^{48}$Ti, $^{76}$Ge$\rightarrow$$^{76}$Se,
$^{82}$Se$\rightarrow$$^{82}$Kr (ground state to ground state).

As can be observed, the perturbative behavior is very good for the
theoretical \nmeds~ of $^{48}$Ca and $^{82}$Se, and less satisfactory
for $^{76}$Ge, in line with the results for the electric-quadrupole
$B(E2)$’s.
The agreement with experiment is very good considering the third-order
\heff, as already pointed out in Ref. \cite{Coraggio24a}.

The outcome of the study of the perturbative behavior of the
calculated \heffs~ and, most importantly of the calculated SM wave
functions, makes us confident that the \heffs~ calculated at
third-order in many-body perturbation theory are the best starting
point to study the \zbb~ decay of $^{48}$Ca, $^{76}$Ge, and $^{82}$Se.
Then, in the following section the calculation of \nmes~ will be
performed by way of third-order \heffs~ to concentrate the attention
to the perturbative properties of the SM effective \zbb-decay
operator, and the related uncertainty estimate of the nuclear matrix
elements \nmes.

\subsection{Neutrinoless double-$\beta$ decay of $^{48}$Ca, $^{76}$Ge,
  and $^{82}$Se}\label{nmeLO}

As introduced in Sec. \ref{theoryho}, in the present work the calculation
of \nme~ accounts for the light-neutrino exchange mechanism, the total
nuclear matrix element being expressed as in Eq. (\ref{nmeLS}) and
calculated accordingly to Eqs. (\ref{operatorGT},\ref{operatorF},\ref{operatorT},\ref{operatorCT},\ref{M0nuappl},\ref{M0nuappCT},\ref{neutpotappl},\ref{neutpotappCT}).

As in our previous studies of the \zbb~ decay
\cite{Coraggio20a,Coraggio22a}, we have carried out the perturbative
expansion of the \zbb~ effective operator \thetaeff~ including in the
\tbox~ diagrams up to the third order (see Section \ref{effhop}), and
a number of intermediate states which corresponds to oscillator quanta
up to $N_{\rm  max}=18$.
In fact, this value corresponds to a number of intermediate states
that is large enough to guarantee a substantial convergence of the
calculated \nmes.

\begin{figure}[h]
\begin{center}
\includegraphics[scale=0.32,angle=0]{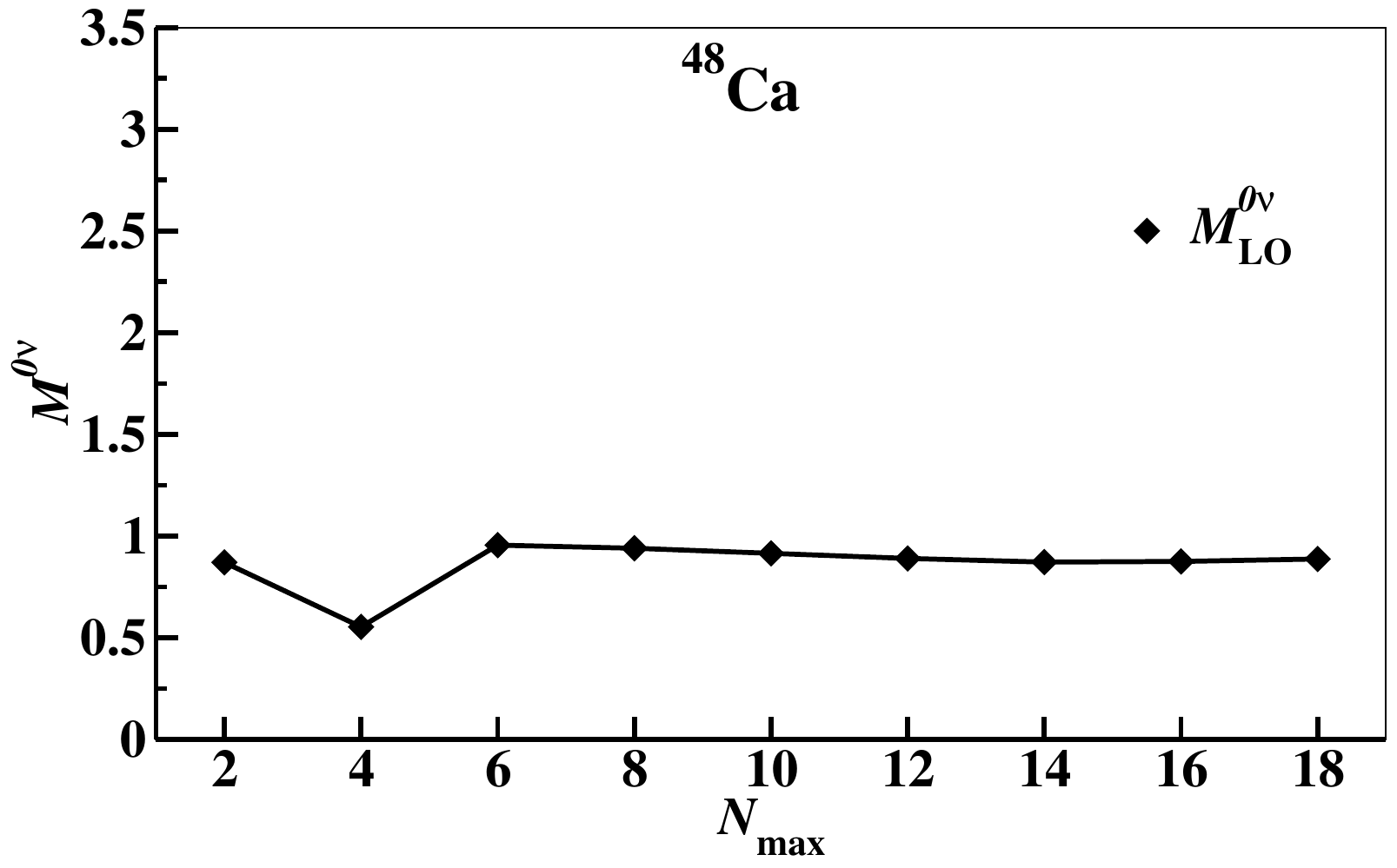}
\caption{\nme~ for the $^{48}\mbox{Ca} \rightarrow ^{48}$Ti decay as a
  function of $N_{\rm max}$}
\label{figNmax}
\end{center}
\end{figure}

In Fig. \ref{figNmax} the calculated values of \nme~for the ${\rm
  ^{48}Ca} \rightarrow {\rm ^{48}Ti}$ decay are reported as a function
of $N_{\rm max}$, and, as can be observed, results are substantially
convergent from $N_{\rm max}=14$ on.

This convergent behavior appears also for the \nmes~ calculated for
the $^{76}$Ge and $^{82}$Se decays, as reported in Table
\ref{ME_0nbb}.

\begin{table}[ht]
  \caption{\nmes~ for $^{76}$Ge and $^{82}$Se \zbb~ decay between
    their ground states, calculated for $N_{\rm max}=14$, 16, and 18.}
\begin{ruledtabular}
\begin{tabular}{cccc}
  \label{ME_0nbb}
  Decay & $N_{\rm max}=14$ & $N_{\rm max}=16$ & $N_{\rm max}=18$ \\
  ~ & ~ & ~ & ~ \\
$^{76}$Ge$\rightarrow$$^{76}$Se & 1.698 & 1.658 & 1.661 \\
 $^{82}$Se$\rightarrow$$^{82}$Kr & 1.284 & 1.250 & 1.253 \\
\end{tabular}
\end{ruledtabular}
\end{table}

Now, we shift the focus on the results of the order-by-order
convergence, that are the starting point for discussing the
uncertainties related to our calculated \nmes.

\begin{figure}[h]
\begin{center}
\includegraphics[scale=0.34,angle=0]{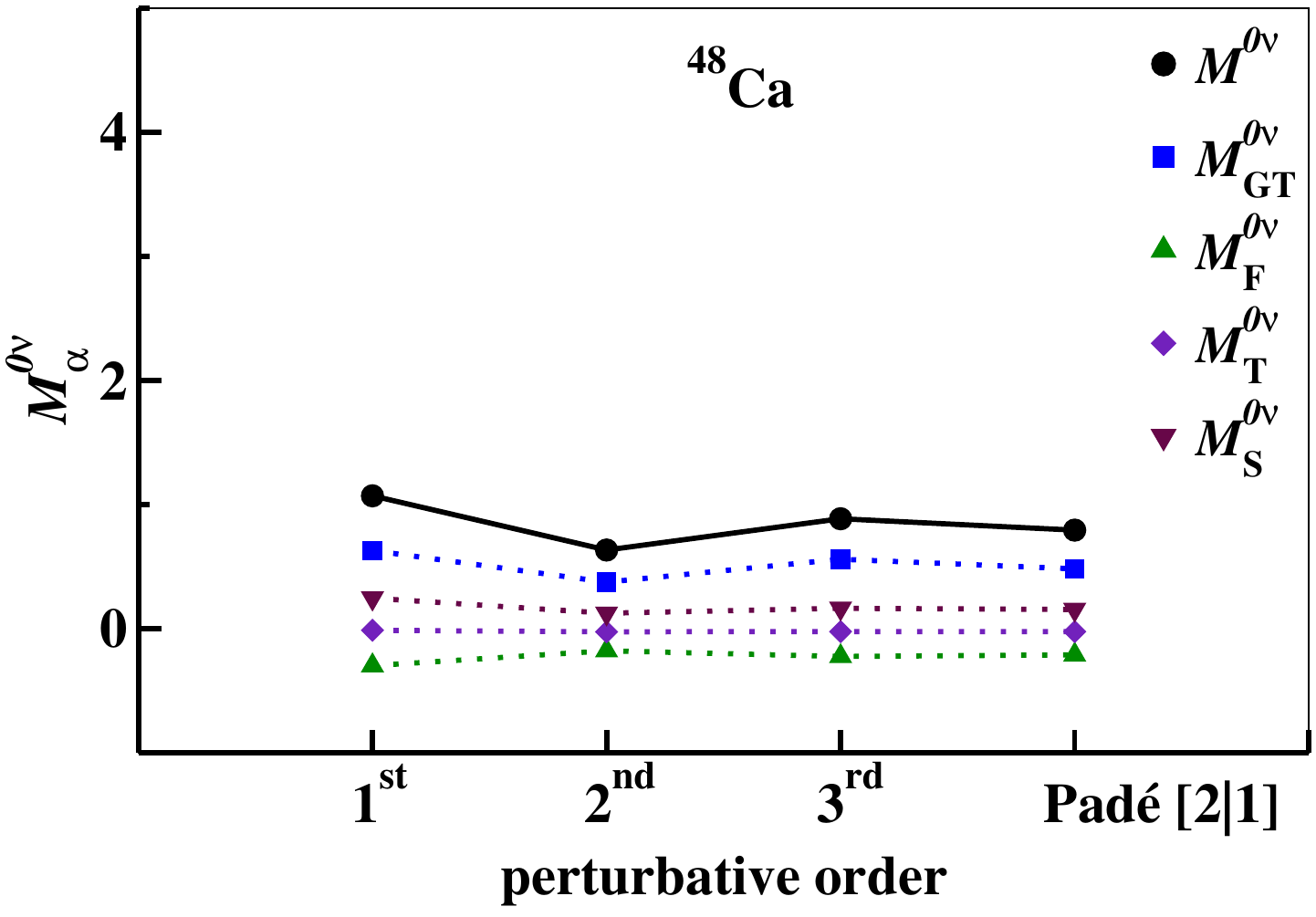}
\caption{\nme~ for the decay of the $^{48}$Ca ground state to the
  $^{48}$Ti one, as a function of the perturbative order. 
The green triangles correspond to $M^{0\nu}_{\rm F}$, the blue squares
to $M^{0\nu}_{\rm GT}$, the purple diamonds to $M^{0\nu}_{\rm T}$, the
brown lower triangles to $M^{0\nu}_{\rm S}$, and the black dots to
the full \nme.}
\label{48Ca_obo}
\end{center}
\end{figure}

First, we consider the \zbb~ decay between the ground states of
$^{48}$Ca and $^{48}$Ti.
In Fig. \ref{48Ca_obo} we have reported the calculated values of \nme,
$M^{0\nu}_{\rm GT}$, $M^{0\nu}_{\rm F}$, $M^{0\nu}_{\rm T}$, and
$M^{0\nu}_{\rm S}$ from first- up to third-order in perturbation
theory.
We have decided to report also the value of their Pad\'e approximant
$[2|1]$, as an indicator of the quality of the perturbative behavior
of the calculated \nme~ \cite{Baker70}.

As can be seen in Fig. \ref{48Ca_obo}, the perturbative behavior is driven
by the Gamow-Teller component, since LO contact term is rather flat
between second- and third-order, and both Fermi and tensor matrix
elements $M^{0\nu}_{\rm F}$, $M^{0\nu}_{\rm T}$ are weakly affected by
the renormalization procedure.

\begin{figure}[h]
\begin{center}
\includegraphics[scale=0.34,angle=0]{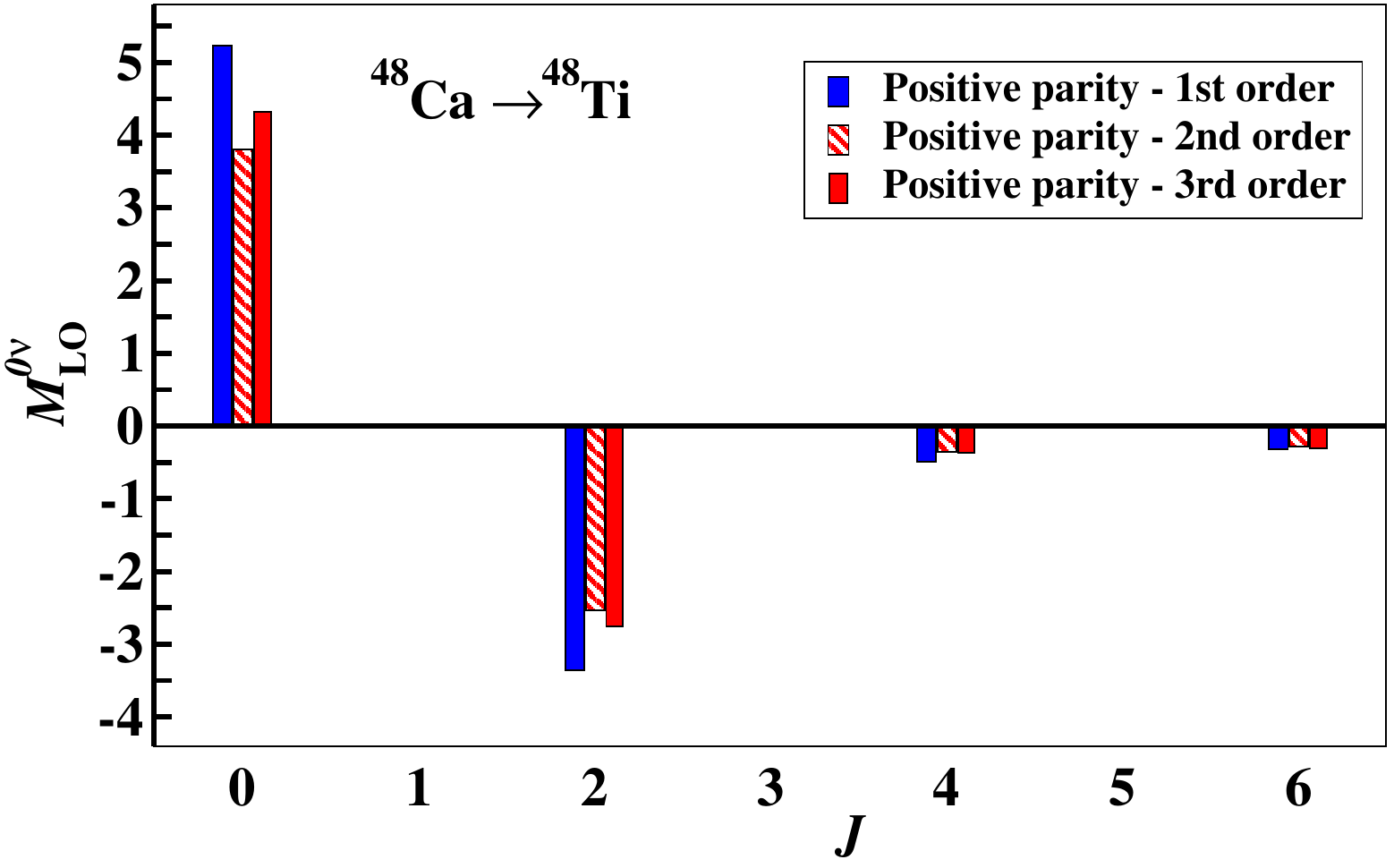}
\caption{Contributions from pairs of decaying neutrons with given
  $J^{\pi}$ to $M^{0\nu}$ for $^{48}$Ca \zbb~decay. The bars
  filled in blue, dashed red, and filled in red correspond to the
  results obtained with \thetaeff~ calculated at first-, second-, and
  third-order in perturbation theory, respectively.}
\label{48Ca-jj}
\end{center}
\end{figure}

A more refined analysis of our results and of the perturbative
behavior may be obtained performing a decomposition of \nmes~ in terms
of the contributions from the decaying pair of neutrons coupled to a
given angular momentum and parity $J^{\pi}$.
This decomposition for \zbb~ decay of $^{48}$Ca is reported in
Fig. \ref{48Ca-jj}, comparing the contributions obtained by employing
the SM effective \zbb-decay operator \thetaeff~ as calculated at first-,
second-, and third-order in many-body perturbation theory.

There are two remarks that we point out from the inspection of
Fig. \ref{48Ca-jj}: first, the main contributions, at all orders, are
provided by $J^{\pi}=0^+,2^+$ components and they are always of
opposite sign, and this fact is mainly responsible for the
particularly small \nme~ for the decay of $^{48}$Ca, a feature that is
common to almost all calculations for this doubly-closed nuclear
system \cite{Senkov13,Senkov16,Senkov14,Jiao18,Coraggio20a}.

Second, we note that the perturbative behavior of each component is
much better than the total \nme, as observed in Fig. \ref{48Ca_obo},
and that it is just the cancellation between the two main components
which contributes largely to the increase of the uncertainty of the
calculated \nme, with respect to the one that characterizes each term
of the decomposition.

\begin{figure}[h]
\begin{center}
\includegraphics[scale=0.34,angle=0]{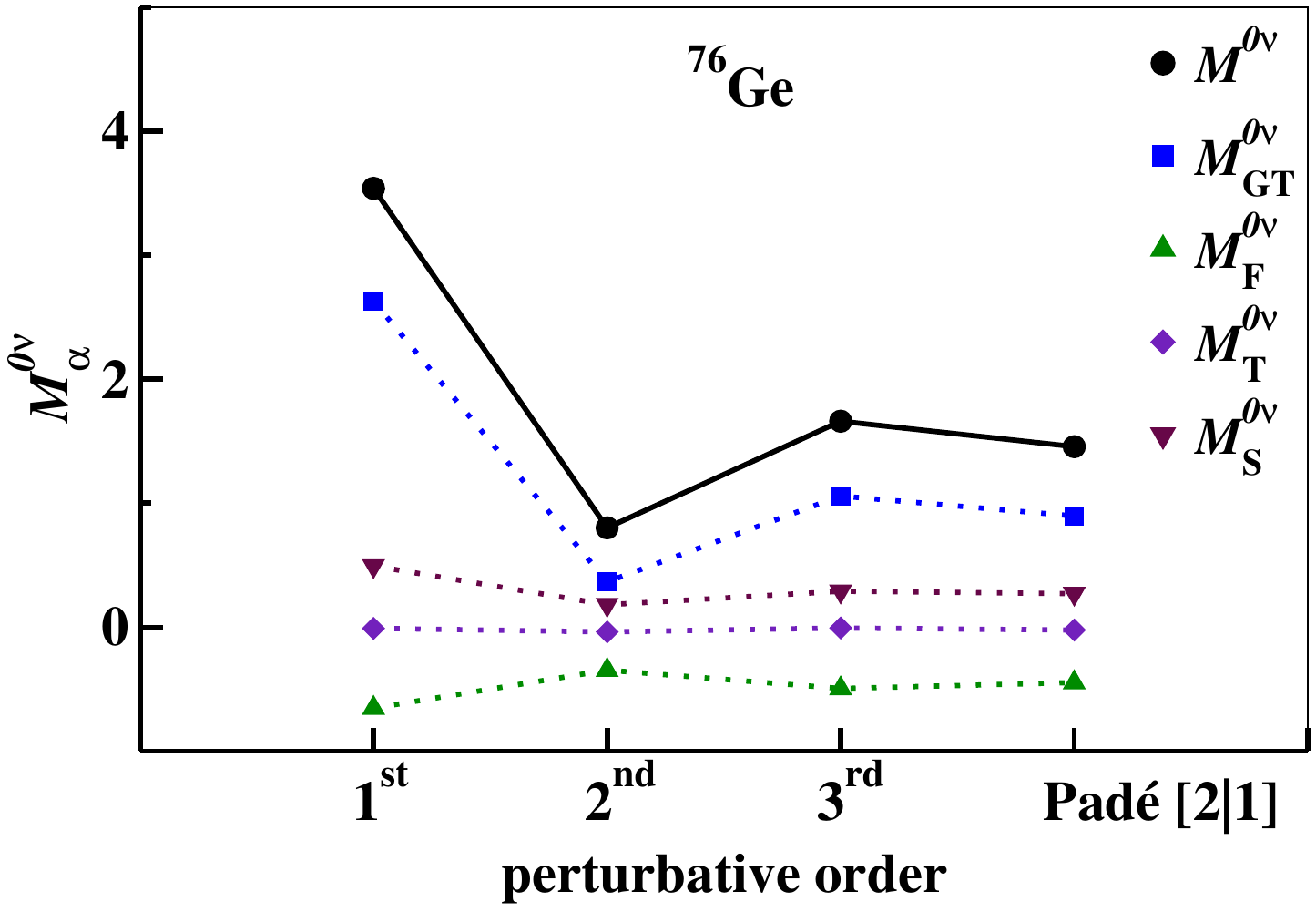}
\caption{Same as in Fig. \ref{48Ca_obo}, but for the decay of the
  $^{76}$Ge ground state to the $^{76}$Se one.}
\label{76Ge_obo}
\end{center}
\end{figure}

\begin{figure}[h]
\begin{center}
\includegraphics[scale=0.34,angle=0]{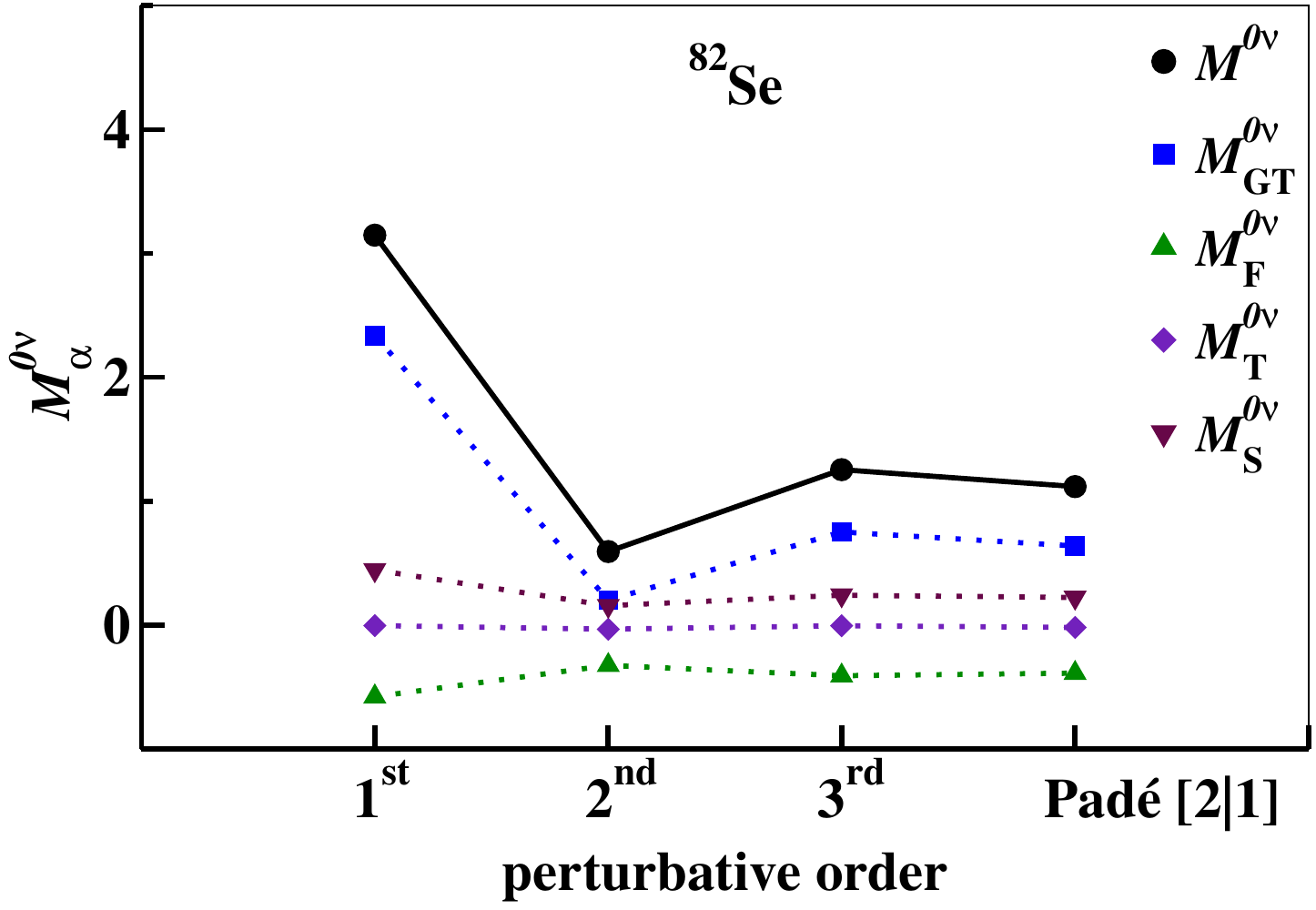}
\caption{Same as in Fig. \ref{48Ca_obo}, but for the decay of the
  $^{82}$Se ground state to the $^{82}$Kr one.}
\label{82Se_obo}
\end{center}
\end{figure}

Very similar observations can be drawn also for the \zbb~ decay of the
ground states of $^{76}$Ge, $^{82}$Se to the ones of $^{76}$Se,
$^{82}$Kr, from the inspection of Figs. \ref{76Ge_obo},\ref{82Se_obo}
which report the calculated values of \nme, $M^{0\nu}_{\rm
  GT}$, $M^{0\nu}_{\rm F}$, $M^{0\nu}_{\rm T}$, and $M^{0\nu}_{\rm S}$
from first- up to third-order in perturbation theory, as well as the
value of their Pad\'e approximant $[2|1]$.

For these decays too, the decomposition of \nmes~ in terms of the
contributions of the decaying pair of neutrons coupled to a given
angular momentum and parity $J^{\pi}$ provides a useful insight on the
main sources of the uncertainties related to the perturbative
expansion of \zbb~ \thetaeffs.

\begin{figure}[h]
\begin{center}
\includegraphics[scale=0.44,angle=0]{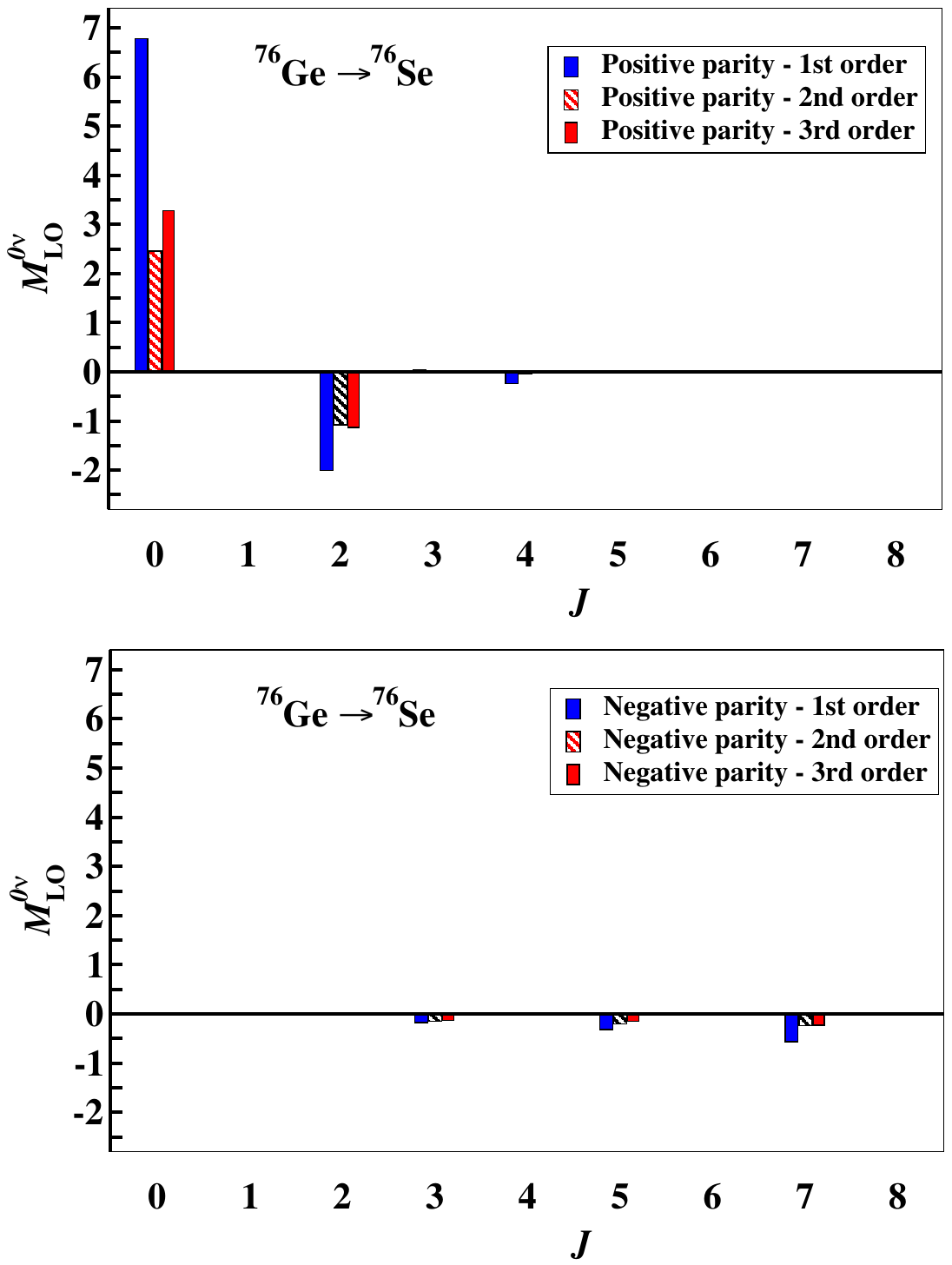}
\caption{ Same as in Fig. \ref{48Ca-jj}, but for the decay of the
  $^{76}$Ge ground state to the $^{76}$Se one.}
\label{76Ge-jj}
\end{center}
\end{figure}

\begin{figure}[h]
\begin{center}
\includegraphics[scale=0.44,angle=0]{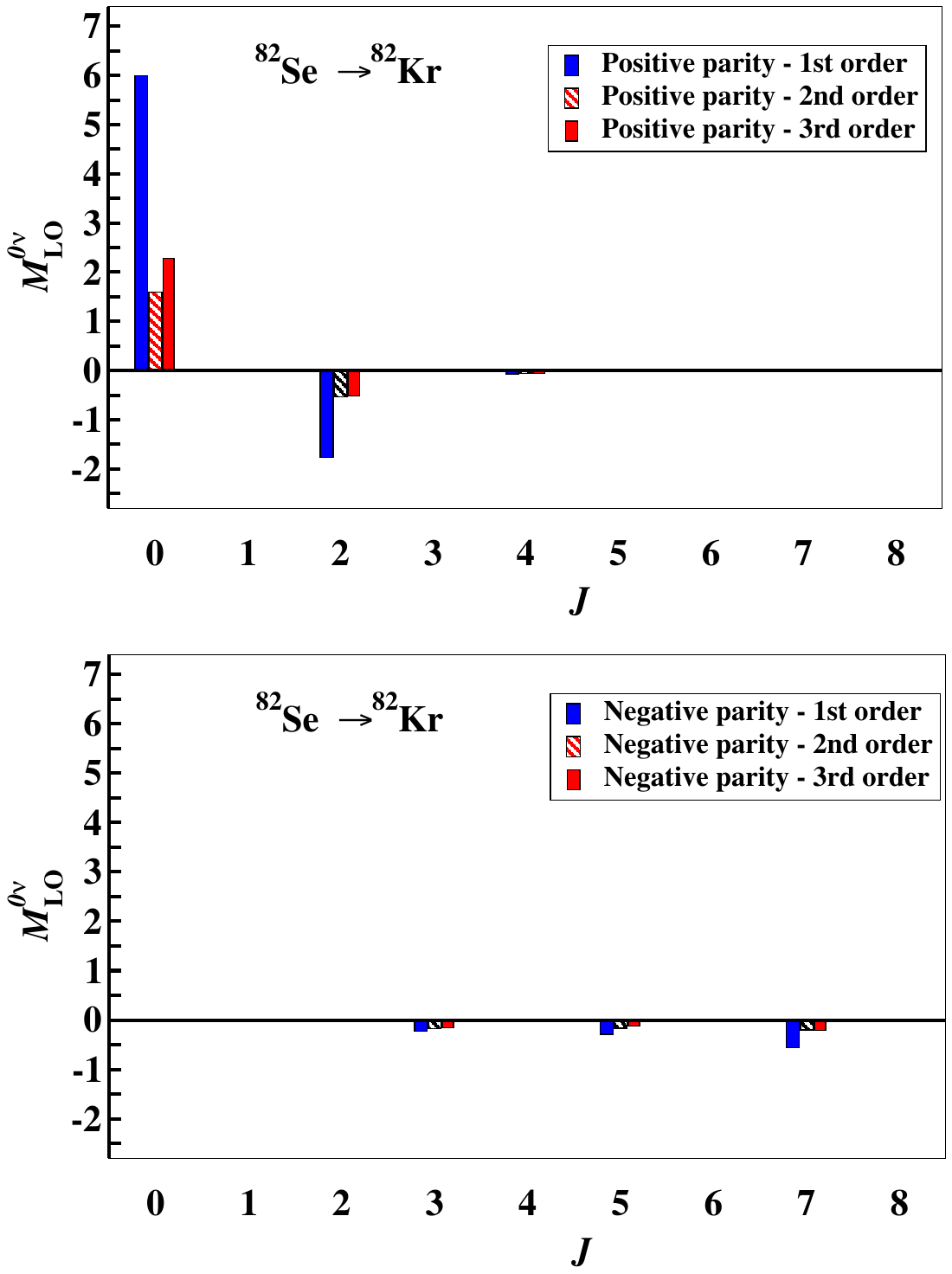}
\caption{ Same as in Fig. \ref{48Ca-jj}, but for the decay of the
  $^{82}$Se ground state to the $^{82}$Kr one.}
\label{82Se-jj}
\end{center}
\end{figure}

From the inspection of Figs. \ref{76Ge-jj},\ref{82Se-jj}, we observe
that the most relevant contributions come by the $J^{\pi}=0^+,2^+$
components and they are always of opposite sign, as for the decay of
$^{48}$Ca, but their respective intensities differ more substantially, and
lead to larger \nmes~ with respect to the $^{48}$Ca decay.

The perturbative behavior is dominated by the $J^{\pi}=0^+$ component,
that is characterized by a larger difference between second- and
third-order calculations.

This exposition of our results leads to an analysis of our evaluation
of \nmes~ for $^{48}$Ca, $^{76}$Ge, and $^{82}$Se \zbb~ decays, as
well as an estimate of the corresponding theoretical uncertainties.

It should be stressed that a calculation that is grounded on many-body
perturbation theory is not a size-extensive approach as for {\it ab
  initio} methods, then a proper theoretical error cannot be
evaluated.
Then, we rely on the theory of Pad\'e approximant as an instrument to
obtain the best approximation of the sum of a perturbative expansion
from a truncated power series \cite{Baker70}.
Moreover, here we refer only to the many-body perturbation theory as a
source of uncertainties, while a complete treatment should involve
also a study of the perturbativity of the ChPT expansion of the
nuclear Hamiltonian and electroweak currents, the latter defining the
decay operators.

\begin{table}[H]
  \caption{Calculated values of \nme~ for all decays under
    investigation. The first column corresponds to the results
    obtained employing the SM effective \zbb-decay operator at
    third-order in perturbation theory, the second one is the
    corresponding Pad\'e approximant $[2|1]$ . In the third column the
    absolute value of the difference of results in columns one and two
    ($\Delta$) are reported.}
\begin{ruledtabular}
\begin{tabular}{cccc}
\label{NME}
 Decay & $M^{0\nu}_{\rm 3rd}$ & $M^{0\nu}_{\textrm {Pad\'e}}$ & $\Delta$ \\
\colrule
~ & ~ & ~& ~ \\
$^{48}$Ca  $\rightarrow$ $^{48}$Ti & 0.89 & 0.80 & 0.09 \\
$^{76}$Ge  $\rightarrow$ $^{76}$Se & 1.66 & 1.46 & 0.20 \\
$^{82}$Se  $\rightarrow$ $^{82}$Kr & 1.25 & 1.12 & 0.13 \\
\colrule
\end{tabular}
\end{ruledtabular}
\end{table}

On the above grounds, we have reported in Table \ref{NME} the values
of \nmes~ as obtained at third order in many-body perturbation theory,
and the Pad\'e approximant $[2|1]$ which accounts for the perturbative
behavior up to third-order.
In the last column, we have reported the absolute value of the
difference between these two values $\Delta=|M^{0\nu}_{\rm 3rd} -
M^{0\nu}_{\textrm {Pad\'e}}|$, that we propose as the estimate of the
uncertainties associated to our calculated \nmes.

To support such a choice, the quantities in Table \ref{ME_2nbb2}
provide an additional check of the perturbative behavior of the
effective double-$\beta$ decay operators.
As can be seen, the $M^{2\nu}$ values obtained with the third-order
effective operator (which are also reported in Table \ref{ME_2nbb}) are
very close to the  corresponding Pad\'e $[2|1]$ approximants, with
differences that are negligible on the scale of the matrix elements.
It should be stressed that, although the \dbb~ and \zbb~ decays
involve different transition operators, both calculations rely on the
same many-body perturbative construction of effective shell-model
operators.

Finally, it is worth to compare our results with those reported in
recent papers, where the \zbb~ decay for intermediate- and heavy-mass
systems has been investigated within EFT electroweak operators.

\begin{table}[H]
  \caption{Same as in Table \ref{NME}, but for \nmeds~ and including
    the corresponding experimental values.}
\begin{ruledtabular}
\begin{tabular}{ccccc}
\label{ME_2nbb2}
 Decay & $M^{2\nu}_{\rm 3rd}$ & $M^{2\nu}_{\textrm {Pad\'e}}$ & $\Delta$ & Expt \\
\colrule
~ & ~ & ~& ~ & ~ \\
$^{48}$Ca$\rightarrow$$^{48}$Ti & 0.019 & 0.019 & 0.0 & $0.042 \pm 0.004$ \\
$^{76}$Ge$\rightarrow$$^{76}$Se & 0.118 & 0.114 & 0.004 & $0.129 \pm 0.004$ \\
$^{82}$Se$\rightarrow$$^{82}$Kr & 0.095 & 0.093 & 0.002 & $0.103 \pm 0.001$ \\
\end{tabular}
\end{ruledtabular}
\end{table}

In Ref. \cite{Belley24}, the authors carried out a comprehensive {\it
  ab initio} uncertainty quantification of the \zbb~ decay of $^{76}$Ge,
employing nuclear Hamiltonians and electroweak operators derived
within ChPT.
$^{76}$Ge \nme~ was calculated with recently developed many-body
emulators, and their numerical result is $M^{0\nu} =
2.60^{+1.28}_{-1.36}$.

The latter is greater than the value reported in Table \ref{NME} --
$M^{0\nu} =1.46$ --, but both \nmes~ are consistent within the
theoretical error in Ref. \cite{Belley24} and our estimated
uncertainty.

The authors of the study in Ref. \cite{Castillo25} evaluated the
\nmes~ for several \zbb~ decays, employing \zbb~ decay operators
derived up to N$^2$LO in ChPT, and employing both pnQRPA and
nuclear shell model, but with empirical effective Hamiltonians.
The SM results reported there in Table 1 evidence values for \nmes~ of
$^{48}$Ca, $^{76}$Ge, and $^{82}$Se, for the LO long- and short-range
components of the \zbb~ decay operator, larger than those reported
in Table \ref{NME}.

Note that with empirical \heffs, as in Ref. \cite{Castillo25}, it is
not possible to construct consistent SM effective decay operators. 
Therefore, their results need to be compared with our results obtained
with our \zbb~ \thetaeff~ at first order in many-body perturbation
theory.

For $^{48}$Ca decay, our first-order long-range $M^{0\nu}_{\rm
  L}=0.83$, the short-range component being $M^{0\nu}_{\rm S}=0.25$ at
first order.
Considering $^{76}$Ge and $^{82}$Se decays, our first-order values are
$M^{0\nu}_{\rm L}=3.05,2.70$ and $M^{0\nu}_{\rm S}=0.49,0.44$,
respectively, which are in a much closer agreement with the central
values of Ref. \cite{Castillo25}, considering the ranges reported
there in Table 1.

\section{Summary and Outlook}
\label{conclusions}
In this work, for the first time, we have carried out a shell-model
calculation of the matrix elements for the \zbb-decay $^{48}$Ca,
$^{76}$Ge, and $^{82}$Se, employing consistent effective Hamiltonians
and decay operators derived within the ChPT.
In particular, we have started from the nuclear Hamiltonian which has
been constructed including two-body contributions up to N$^3$LO and
three-body ones up to N$^2$LO, and the \zbb-decay operator included
only the leading-order (LO) contribution for the light-neutrino
exchange.

The derivation of the SM effective operators has been performed
through many-body perturbation theory, including all contributions up
to third order.
The reliability of such an approach has been tested in a previous work
\cite{Coraggio24a}, comparing spectroscopic observables and
experimental GT matrix elements with the theoretical ones.
Here we have also added a discussion about the convergence properties
of calculated low-energy spectra and \dbb-decay matrix elements for
the nuclei under scrutiny, aiming to establish the soundness of the
many-body perturbative expansion.

Next, the focus of the perturbativity was on the calculated \nmes,
since one of our interests has been to estimate the uncertainties
associated to them because of the perturbative approach.
Since we employ a many-body perturbative approach to carry out the
shell-model calculations, the evaluation of the uncertainties related
to \nmes~ has been grounded on the theory of Pad\'e approximants, and
does not account for the theoretical errors that are associated to the
construction of the nuclear Hamiltonian and electroweak decay
operators within the chiral perturbation theory, and that is also
tackled in other recent works \cite{Belley24}.

The outlook of our study of calculating \nmes – and in general
electroweak decay observables that may be related to such a rare
process – is to shift our efforts in considering higher-order
contributions of the ChPT expansion of the \zbb~ currents, namely
including in the SM effective decay operator terms up to N$^2$LO, and
then investigating also the perturbative properties of approaching the
\zbb~ decay within EFT.

Moreover, we plan to apply this framework to study also the
double-$\beta$ decay of $^{100}$Mo, a nuclear system that is currently
a candidate to the detection of the \zbb~ decay, but that is very
challenging from the point of view of a shell-model calculation.

\section{Acknowledgments}
The calculations were made possible by grants of computing time from
the Italian National Supercomputing Center CINECA, under the
CINECA-INFN agreement.

\bibliographystyle{apsrev}
\bibliography{biblio.bib}

\end{document}